\documentclass[10pt, twocolumn, journal]{IEEEtran}
\usepackage{graphics, subfigure, color, epsfig, psfrag, amsmath, amsfonts, amssymb, eucal, balance}
\newtheorem{thm}{Theorem}
\newtheorem{cor}{Corollary}

\newtheorem{prop}{Proposition}
\newtheorem{defn}{Definition}

\DeclareMathAlphabet{\eurm}{U}{eur}{m}{n}
\DeclareMathAlphabet{\mathbsf}{OT1}{cmss}{bx}{n}
\DeclareMathAlphabet{\mathssf}{OT1}{cmss}{m}{sl}
\DeclareMathAlphabet{\mathcsf}{OT1}{cmss}{sbc}{n}

\newcommand{\randomvalue}[1]{\eurm{\uppercase{#1}}}

\DeclareSymbolFont{bsfletters}{OT1}{cmss}{bx}{n}  
\DeclareSymbolFont{ssfletters}{OT1}{cmss}{m}{n}
\DeclareMathSymbol{\bsfGamma}{0}{bsfletters}{'000}
\DeclareMathSymbol{\ssfGamma}{0}{ssfletters}{'000}
\DeclareMathSymbol{\bsfDelta}{0}{bsfletters}{'001}
\DeclareMathSymbol{\ssfDelta}{0}{ssfletters}{'001}
\DeclareMathSymbol{\bsfTheta}{0}{bsfletters}{'002}
\DeclareMathSymbol{\ssfTheta}{0}{ssfletters}{'002}
\DeclareMathSymbol{\bsfLambda}{0}{bsfletters}{'003}
\DeclareMathSymbol{\ssfLambda}{0}{ssfletters}{'003}
\DeclareMathSymbol{\bsfXi}{0}{bsfletters}{'004}
\DeclareMathSymbol{\ssfXi}{0}{ssfletters}{'004}
\DeclareMathSymbol{\bsfPi}{0}{bsfletters}{'005}
\DeclareMathSymbol{\ssfPi}{0}{ssfletters}{'005}
\DeclareMathSymbol{\bsfSigma}{0}{bsfletters}{'006}
\DeclareMathSymbol{\ssfSigma}{0}{ssfletters}{'006}
\DeclareMathSymbol{\bsfUpsilon}{0}{bsfletters}{'007}
\DeclareMathSymbol{\ssfUpsilon}{0}{ssfletters}{'007}
\DeclareMathSymbol{\bsfPhi}{0}{bsfletters}{'010}
\DeclareMathSymbol{\ssfPhi}{0}{ssfletters}{'010}
\DeclareMathSymbol{\bsfPsi}{0}{bsfletters}{'011}
\DeclareMathSymbol{\ssfPsi}{0}{ssfletters}{'011}
\DeclareMathSymbol{\bsfOmega}{0}{bsfletters}{'012}
\DeclareMathSymbol{\ssfOmega}{0}{ssfletters}{'012}

\newcommand{\rvl}{{\randomvalue{l}}}    

\newcommand{\rvn}{{\randomvalue{n}}}	

\newcommand{\rvt}{{\randomvalue{t}}}	

\newcommand{\rvx}{{\randomvalue{x}}}	

\newcommand{\rvz}{{\randomvalue{z}}}	

\begin{document}
\title{Opportunistic Detection Rules: Finite and Asymptotic Analysis}

\author{Wenyi Zhang, George V. Moustakides, and H. Vincent Poor
\thanks{W. Zhang is with Department of Electronic Engineering and Information Sciences, University of Science and Technology of China, China (Email: wenyizha@ustc.edu.cn); G. V. Moustakides is with Department of Electrical and Computer Engineering, University of Patras, Greece (Email: moustaki@upatras.gr); H. V. Poor is with Department of Electrical Engineering, Princeton University, USA (Email: poor@princeton.edu). The work of W. Zhang was supported by National Natural Science Foundation of China under grant 61379003, the National Basic Research Program of China (973 Program) through grant 2012CB316004, and the SRFDP-RGC ERG Joint Research Scheme through Specialized Research Fund 20133402140001; the work of H. V. Poor was supported by the U. S. National Science Foundation under Grants DMS-1118605 and ECCS-1343210.}
}

\maketitle

\begin{abstract}
Opportunistic detection rules (ODRs) are variants of fixed-sample-size detection rules in which the statistician is allowed to make an early decision on the alternative hypothesis opportunistically based on the sequentially observed samples. From a sequential decision perspective, ODRs are also mixtures of one-sided and truncated sequential detection rules. Several results regarding ODRs are established in this paper. In the finite regime, the maximum sample size is modeled either as a fixed finite number, or a geometric random variable with a fixed finite mean. For both cases, the corresponding Bayesian formulations are investigated. The former case is a slight variation of the well-known finite-length sequential hypothesis testing procedure in the literature, whereas the latter case is new, for which the Bayesian optimal ODR is shown to be a sequence of likelihood ratio threshold tests with two different thresholds: a running threshold, which is determined by solving a stationary state equation, is used when future samples are still available, and a terminal threshold (simply the ratio between the priors scaled by costs) is used when the statistician reaches the final sample and thus has to make a decision immediately. In the asymptotic regime, the tradeoff among the exponents of the (false alarm and miss) error probabilities and the normalized expected stopping time under the alternative hypothesis is completely characterized and proved to be tight, via an information-theoretic argument. Within the tradeoff region, one noteworthy fact is that the performance of the Stein-Chernoff Lemma is attainable by ODRs.
\end{abstract}

\section{Introduction}
\label{sec:intro}

In this paper we consider a setup of discriminating two simple hypotheses, as follows. At most $N$ independent and identically distributed (i.i.d.) random variables (called ``samples'' interchangeably throughout the paper) $\rvx_i$, $i = 1, 2, \ldots, N$, from one of two distributions $p_0$ (the null hypothesis $\mathcal{H}_0$) and $p_1$ (the alternative hypothesis $\mathcal{H}_1$), are drawn sequentially one at a time by a statistician, who is allowed to stop sampling at some time $T < N$ and decide $\mathcal{H}_1$, or wait until observing all the $N$ samples and decide either $\mathcal{H}_0$ or $\mathcal{H}_1$. A strategy adopted by the statistician is called an ``opportunistic detection rule'' (ODR).

To motivate this ODR setup, on one hand, note that it is often imperative to attain a small decision delay under $\mathcal{H}_1$ (which may correspond to an abnormal condition that requires an immediate attention), but of less importance to stop promptly under $\mathcal{H}_0$ (which may correspond to a normal condition),\footnote{Other application scenarios are also possible. For example, consider inspecting a number of products that were made by the same machine, to determine whether they have some common flaw or not. Inspecting a product will destroy it. Therefore, it is desirable to inspect as few products as possible if they do not have the flaw, but immaterial if they do.} on the other hand, the finite maximum sample size is reasonable since in any application it is unlikely to have an infinite number of samples or to spend an infinite amount of time to make a decision.

Such an ODR setup is closely related to both fixed-sample-size (FSS) hypothesis testing and sequential hypothesis testing. Subsequently we briefly discuss the background to place our work into the context.
\subsection{Related Work}
\label{subsec:literature}

\subsubsection{FSS Hypothesis Testing}

Discriminating two distributions based on an array of i.i.d. samples is of fundamental importance in statistical decision theory. Of particular interest, characterizing the asymptotic performance limit, in terms of achievable exponential decay rates of the (false alarm and miss) error probabilities as the sample size grows without bound, has received significant attention in the literature. The Stein-Chernoff Lemma \cite{chernoff56:ams} \cite{stein} characterizes the maximum achievable exponential decay rate of the miss probability for any fixed false alarm probability, under a Neyman-Pearson setting. Under a Bayesian setting which seeks to minimize a linear combination of the false alarm and miss probabilities, the Chernoff information \cite{chernoff52:ams} measures the maximum achievable exponential decay rate of the Bayesian cost. More generally, the optimal tradeoff between the exponential decay rates of the false alarm probability and the miss probability was originally studied in \cite{hoeffding65:ams} and later treated in information theory \cite{csiszar71:ssmh} \cite{blahut74:it}. In the asymptotic analysis of the present paper, we further extend the problem to study the tradeoff among the exponential decay rates of the false alarm probability, the miss probability, and the expected stopping time (normalized by the maximum sample size $N$) under the alternative hypothesis, because here for ODR it is of interest reducing the decision delay under the alternative hypothesis, at the cost of sacrificing the decision reliability.

Beyond the scope of the present paper, later development along this direction includes the strong converse \cite{han89:it} which characterizes the exponential decay rate of the rejection probability (i.e., one minus the false alarm probability) when the exponential decay rate of the miss probability exceeds that indicated by the Stein-Chernoff Lemma, and generalization for discriminating two general (non-i.i.d.) sequences of distributions \cite{han00:it} \cite{nagaoka07:it}.

\subsubsection{Sequential Hypothesis Testing}

When there is no limit on the maximum sample size and the statistician is allowed to wait indefinitely before making decision, the Wald-Wolfowitz Theorem (see, e.g., \cite{wald48:ams} \cite[Sec. 7.6]{ferguson67:book} \cite[Thm. 4.7]{poor09:book}) is fundamental. It asserts that the sequential probability ratio test (SPRT), which sequentially compares the likelihood ratios $\Lambda_k = \prod_{i = 1}^k \frac{p_1(\rvx_i)}{p_0(\rvx_i)}$, $k = 1, 2, \ldots$, against two thresholds $0 < A \leq 1 \leq B < \infty$, and decides $\mathcal{H}_0$ once $\Lambda_k \leq A$ or $\mathcal{H}_1$ once $\Lambda_k \geq B$, is optimal in the sense that, among all possible stopping rules whose false alarm and miss probabilities are no worse than those attained by the SPRT, the SPRT requires the minimum expected stopping times under both hypotheses.\footnote{There is a technical condition that the sum of the false alarm and miss probabilities is no greater than one.}

The requirement in our ODR setup that the statistician stops sampling only when deciding $\mathcal{H}_1$ is closely related to the so-called power-one or one-sided stopping rules, extensively studied in statistics for testing simple and composite hypotheses; see, e.g., \cite{farrell64:ams} \cite{siegmund69:ams} \cite{robbins74:as} \cite{lai77:as}. A common flavor of these earlier works is the emphasis on fine asymptotic behavior of the expected stopping time, usually in form of the law of the iterated logarithm. A simple one-sided stopping rule is the one-sided SPRT, which sequentially compares $\Lambda_k = \prod_{i = 1}^k \frac{p_1(\rvx_i)}{p_0(\rvx_i)}$, $k = 1, 2, \ldots$, against a threshold $1 < B < \infty$, and decide $\mathcal{H}_1$ once $\Lambda_k \geq B$. Note that, the one-sided SPRT may never stop, --- indeed, it stops with probability no greater than $1/B$ under $\mathcal{H}_0$, but stops with probability one under $\mathcal{H}_1$ \cite{fellouris12:sa}. Furthermore, the one-sided SPRT is optimal in the sense that, among all possible stopping rules whose false alarm probabilities are no worse than that attained by the one-sided SPRT, the one-sided SPRT requires the minimum expected stopping time under $\mathcal{H}_1$ \cite[pp. 107-108]{chow71:book}.

The requirement in our ODR setup that the statistician must make a decision upon reaching a finite deadline $N$ essentially defines a truncated sequential hypothesis testing problem. The previous study on truncated stopping rules usually concerns about the performance of composite hypothesis testing with a pair of time-varying threshold sequences that coincide at the truncation point; see, e.g., \cite{anderson60:ams} \cite{bussgang67:it} \cite{tantaratana82:it} \cite{siegmund85:book} and references therein. Meanwhile, truncated stopping rules are often used as an intermediate step when developing stopping rules without a deadline, by passing to the limit of $N \rightarrow \infty$; see, e.g., \cite[5.5]{bertsekas00:book} \cite[3.2.2]{tartakovsky15:book} for the derivation of the SPRT. When combining the one-sided and truncated stopping rules, a specific ODR was recently considered and analyzed in \cite{zhang13:it}: the statistician follows the one-sided SPRT, but decides $\mathcal{H}_0$ if the one-sided SPRT has not stopped before observing the last sample $\rvx_N$. In the finite analysis of the present paper, we develop Bayesian optimal ODRs, for both the case where the maximum sample size is a fixed finite number, and the case where the maximum sample size is a geometric random variable with a fixed finite mean.

In the context of sequential hypothesis testing, asymptotic optimality usually means that the mean (and higher moments, possibly) of the stopping time of a test is close, in a certain asymptotic sense, to that attained by the optimal test with the same error probabilities, which is usually difficult to design analytically or even numerically; see, e.g., \cite{chernoff72:book} \cite{dragalin99:it} \cite{tartakovsky15:book}.\footnote{Relatively few works consider the exponential decay rates of the (false alarm and miss) error probabilities for sequential hypothesis testing. The optimal tradeoff for sequential binary simple hypothesis testing has been characterized in \cite{polyanskiy11:ita}, but the definition of the exponential decay rates therein is different from that considered in FSS hypothesis testing. In the present paper, we follow the definition in FSS hypothesis testing.}

\subsection{Overview of Results}
\label{subsec:overview}

In this paper, we establish several results regarding ODRs.

\subsubsection{Finite Analysis}

In the finite regime, we consider a Bayesian problem formulation, which seeks to characterize the ODR that minimizes a Bayesian cost as a linear combination of the (false alarm and miss) error probabilities and the expected stopping time. We examine two cases, in which the maximum sample size is modeled either as a fixed finite number, or a geometric random variable with a fixed finite mean, respectively. The former case is a slight variation of the well-known finite-length sequential hypothesis testing procedure in the literature (see, e.g., \cite[5.5]{bertsekas00:book} \cite[3.2.2]{tartakovsky15:book}) and is included herein for making the exposition self-contained. The latter case is new. To motivate the model of a random maximum sample size, whose realization is revealed to the statistician only upon observing the last sample, we may consider scenarios in which the observation process is subject to abrupt interruption, or in which an external controller (say a system operator in a smart grid system \cite{hossein12:book}), in a unanticipated manner, issues a command for prompt decision without further observation. For this case, we establish that, interestingly, the Bayesian optimal ODR is a sequence of likelihood ratio threshold tests with two different thresholds: a ``running threshold'', which is determined by solving a stationary state equation, is used when future samples are still available, and a ``terminal threshold'' (simply the ratio between the priors scaled by costs) is used when the statistician reaches the final sample and thus has to make a decision.

\subsubsection{Asymptotic Analysis}

In the asymptotic regime, we let the maximum sample size $N$ grow without bound, and characterize the tradeoff among the exponential decay rates of the (false alarm and miss) error probabilities and the normalized expected stopping time under the alternative hypothesis. As an extreme case in the tradeoff, the asymptotic performance of the optimal fixed-sample-size (FSS) decision rule, described by the Stein-Chernoff Lemma, i.e., an error exponent of $D(p_0\|p_1)$ for the miss probability, is shown to be achievable for any fixed target false alarm probability, with asymptotically vanishing normalized expected stopping time under $\mathcal{H}_1$. The truncated one-sided SPRT ODR considered in \cite{zhang13:it} is thus suboptimal since it achieves an exponent of only $C(p_0, p_1)$, the Chernoff information of $(p_0, p_1)$. When establishing the converse part, i.e., proving that the tradeoff is tight, a key idea of the proof makes use of the converse for the channel capacity per unit cost \cite{verdu90:it} \cite{csiszar73:pcit}.

The remaining part of this paper is organized as follows. Section \ref{sec:bayesian} presents the finite analysis, characterizing the Bayesian optimal ODRs, for both the case of fixed and geometrically distributed maximum sample sizes. Section \ref{sec:tradeoff} presents results of the asymptotic analysis, fully characterizing the tradeoff among the exponents of the false alarm and miss probabilities and the expected stopping time under $\mathcal{H}_1$. Finally, Section \ref{sec:conclusion} concludes this paper.

\section{Finite Analysis}
\label{sec:bayesian}

In this section, we consider the finite regime, in which the maximum sample size is modeled either as a fixed finite number, or a geometric random variable with a fixed finite mean. For both cases, we investigate the corresponding Bayesian formulations.

\subsection{Case of Fixed Maximum Sample Size}
\label{subsec:fixed-length}

In this case, the maximum sample size is a fixed finite number $N$. In order to make use of the optimal stopping theory, it turns out to be convenient to formulate the problem in the following way. Consider all stopping times that stop by $N$, $\mathcal{T}^N$, and terminal decision rules $\mathcal{D}: \mathcal{X}^N \mapsto \{\mathcal{H}_0, \mathcal{H}_1\}$. Note that for ODRs we need to consider only the terminal decision rule, because whenever the stopping time $\rvt < N$ the decision is bound to be $\mathcal{H}_1$. Therefore, an ODR can be defined as follows.
\begin{defn}
\label{defn:ODR}
An opportunistic detection rule (ODR) consists of a stopping time $\rvt \in \mathcal{T}^N$, and a terminal decision rule $D: \mathcal{X}^N \mapsto \{\mathcal{H}_0, \mathcal{H}_1\}$, such that, when $\rvt < N$, the decision is $\mathcal{H}_1$, and when $\rvt = N$, the decision is given by $D(\rvx^N)$. Herein $\rvx^N$ denotes $\rvx_1, \rvx_2, \ldots, \rvx_N$, and thereafter we usually suppress it and simply write $D(\rvx^N)$ as $D$.
\end{defn}

The detection error events can thus be written as
\begin{eqnarray}
\label{eqn:false-alarm-event}
&&\mbox{False alarm:} \{\rvt < N\} \cup \{\rvt = N, D = \mathcal{H}_1\} \;\mbox{w.r.t.} p_0\\
&&\mbox{Miss:} \{\rvt = N, D = \mathcal{H}_0\} \;\mbox{w.r.t.} p_1,
\end{eqnarray}
and the expected stopping time under $\mathcal{H}_1$ is simply $T = \mathbb{E}_1[\rvt]$.

We thus formulate the Bayesian cost as follows:
\begin{eqnarray}
\label{eqn:bayesian-risk}
\mathcal{J} &=& (1 - \pi) c_0 P_\mathrm{FA} + \pi c_1 P_\mathrm{M} + c T\nonumber\\
&=& (1 - \pi) c_0 \mathbb{E}_0\left[\mathbf{1}(\rvt < N) + \mathbf{1}(\rvt = N) \mathbf{1}(D = \mathcal{H}_1)\right] \nonumber\\
&&\quad\quad + \pi c_1 \mathbb{E}_1\left[\mathbf{1}(\rvt = N) \mathbf{1}(D = \mathcal{H}_0)\right] + c \mathbb{E}_1[\rvt]
\end{eqnarray}
where $0 \leq \pi \leq 1$ is the prior probability of hypothesis $\mathcal{H}_1$, and $c_0, c_1, c > 0$ are cost assignments. The problem we seek to solve is then to choose a stopping time $\rvt$ and a terminal decision rule $D$ that minimize $\mathcal{J}$, i.e.,
\begin{eqnarray}
\label{eqn:bayesian-form}
\min_{\rvt \in \mathcal{T}^N, D} \mathcal{J}.
\end{eqnarray}

The problem (\ref{eqn:bayesian-form}) is a slight variation of the well-known finite-length sequential hypothesis testing problem, which is often used as an intermediate step when developing stopping rules without a maximum sample size constraint; see, e.g., \cite[5.5]{bertsekas00:book} \cite[3.2.2]{tartakovsky15:book}. The main difference is that for the ODR problem (\ref{eqn:bayesian-form}), the decision is one-sided so that only a terminal decision rule upon observing the last sample is needed and that the false alarm event (\ref{eqn:false-alarm-event}) is different from that considered in the literature. On the other hand, the derivation of the one-sided SPRT in \cite[pp. 107-108]{chow71:book} directly works with the non-truncated case so that it is not applicable to (\ref{eqn:bayesian-form}) here. Having formulated (\ref{eqn:bayesian-form}), the remaining analysis is a standard exercise of Markov optimal stopping theory, and we include the solution in the remainder of this subsection for the sake of making the exposition self-contained.

We note the following relationship through change of probability measure that transforms $\mathbb{E}_1[\cdot]$ into $\mathbb{E}_0[\cdot]$:
\begin{eqnarray}
\label{eqn:com-m}
&&\mathbb{E}_1\left[\mathbf{1}(\rvt = N) \mathbf{1}(D = \mathcal{H}_0)\right] \nonumber\\
&=& \mathbb{E}_0\left[\Lambda_N \mathbf{1}(\rvt = N) \mathbf{1}(D = \mathcal{H}_0)\right],
\end{eqnarray}
because the miss event $\{\rvt = N, D = \mathcal{H}_0\}$ is $\mathcal{F}_N$-measurable. On the other hand,
\begin{eqnarray}
\label{eqn:com-t}
\mathbb{E}_1[\rvt] &=& \mathbb{E}_1\left[\sum_{k = 1}^N \mathbf{1}(\rvt \geq k)\right]\nonumber\\
&=& \sum_{k = 1}^N \mathbb{E}_1\left[\mathbf{1}(\rvt \geq k) \right]\nonumber\\
&\stackrel{(a)}{=}& \sum_{k = 1}^N \mathbb{E}_0\left[\Lambda_{k - 1} \mathbf{1}(\rvt \geq k)\right]\nonumber\\
&=& \mathbb{E}_0\left[\sum_{k = 1}^N \Lambda_{k - 1} \mathbf{1}(\rvt \geq k)\right]\nonumber\\
&\stackrel{(b)}{=}& \mathbb{E}_0\left[\sum_{k = 0}^{\rvt - 1} \Lambda_k\right],
\end{eqnarray}
where (a) is due to that $\{\rvt \geq k\}$ is $\mathcal{F}_{k - 1}$-measurable, and (b) is due to that $\mathbf{1}(\rvt \geq k) = 0$ for all $k$'s greater than $\rvt$. So with (\ref{eqn:com-m}) and (\ref{eqn:com-t}), we rewrite the Bayesian cost $\mathcal{J}$ as
\begin{eqnarray}
\label{eqn:bayesian-cost-E0}
\mathcal{J} &=& \mathbb{E}_0\left[(1 - \pi)c_0 \left[\mathbf{1}(\rvt < N) + \mathbf{1}(\rvt = N) \mathbf{1}(D = \mathcal{H}_1)\right] +\right.\nonumber\\
 &&\left.\pi c_1 \Lambda_N \mathbf{1}(\rvt = N) \mathbf{1}(D = \mathcal{H}_0) + c \sum_{k = 0}^{\rvt - 1} \Lambda_k\right].
\end{eqnarray}

Inspecting (\ref{eqn:bayesian-cost-E0}), it is clear that for any stopping time $\rvt$, the optimal terminal decision rule is
\begin{eqnarray}
\label{eqn:optimal-D}
&&D = \mathcal{H}_1 \;\;\mbox{if}\; \Lambda_N \geq \frac{(1 - \pi) c_0}{\pi c_1};\nonumber\\
\mbox{and}\;&& D = \mathcal{H}_0 \;\;\mbox{otherwise}.
\end{eqnarray}
This is a quite reasonable result in retrospect, since it coincides with the FSS Bayesian optimal decision rule, which should be the case when the statistician has already observed the last sample.

With the optimal terminal decision rule above, we further rewrite the Bayesian cost $\mathcal{J}$ as
\begin{eqnarray}
\label{eqn:bayesian-cost-T}
&&\mathcal{J} = \mathbb{E}_0\left[c \sum_{k = 0}^{\rvt - 1} \Lambda_k + (1 - \pi)c_0 \mathbf{1}(\rvt < N) + \right.\nonumber\\
&&\left. \mathbf{1}(\rvt = N) \min\{(1 - \pi) c_0, \pi c_1 \Lambda_N\}
\right].
\end{eqnarray}

Now let us characterize the stopping time $\rvt \in \mathcal{T}^N$ that minimizes $\mathcal{J}$. For (\ref{eqn:bayesian-cost-T}), using backward induction (see, e.g., \cite[3.3.1]{poor09:book}), we find that the optimal stopping time is given by
\begin{eqnarray}
\rvt = \min\left\{1 \leq k \leq N - 1: h_k(\Lambda_k) = (1 - \pi) c_0 \right\},
\end{eqnarray}
and if no such $\rvt$ exists, we set $\rvt = N$. The functions $\{h_k\}$ satisfy backward recursion as
\begin{eqnarray}
\label{eqn:backward-recursion}
h_{k - 1}(\lambda) = \min\{(1 - \pi) c_0, c \lambda + \mathbb{E}_0[h_k(\lambda \rvl)]\},
\end{eqnarray}
$k = N, N - 1, \ldots, 2$, where $\rvl = p_1(\rvx)/p_0(\rvx)$ with $\rvx$ obeying $p_0$, and $h_N(\lambda) = \min\{(1 - \pi) c_0, \pi c_1 \lambda\}$.

From the backward recursion (\ref{eqn:backward-recursion}), we have that the optimal ODR is indeed a sequence of likelihood ratio threshold tests.

\begin{thm}
\label{thm:bayesian-structure}
The Bayesian optimal ODR that solves (\ref{eqn:bayesian-form}) is a sequence of likelihood ratio threshold tests, with time-varying thresholds. The thresholds $\tau_k$ are the solutions of
\begin{eqnarray}
\label{eqn:threshold}
c \lambda + \mathbb{E}_0[h_{k + 1} (\lambda \rvl)] = (1 - \pi) c_0,
\end{eqnarray}
for each $k = 1, 2, \ldots, N - 1$, and $\tau_N = \frac{(1 - \pi) c_0}{\pi c_1}$.
\end{thm}

{\it Proof:} Theorem \ref{thm:bayesian-structure} follows from the observations below.

(i) For every $k$, $h_k(\lambda)$ is monotonically non-decreasing with respect to $\lambda > 0$, and is positive except at $\lambda = 0$. This can be directly verified by induction.

(ii) For every $k$, $h_k(\lambda)$ is concave and continuous with respect to $\lambda > 0$. This can be shown by noting that expectation and point-wise minimum operations conserve concavity.

(iii) For every $k$, $h_k(\lambda)$ is equal to $(1 - \pi) c_0$ for any $\lambda$ no smaller than a certain threshold $\tau_k > 0$, and for any $\lambda$ smaller than $\tau_k$, $h_k(\lambda)$ is smaller than $(1 - \pi) c_0$. This is equivalent to the property that the curve $c \lambda + \mathbb{E}_0[h_{k + 1}(\lambda \rvl)]$ crosses the level $(1 - \pi) c_0$ only once, and thus can be shown by using (i), (ii), and the fact that $c \lambda + \mathbb{E}_0[h_{k + 1}(\lambda \rvl)] \rightarrow \infty$ as $\lambda \rightarrow \infty$, for any $c > 0$.

So from (iii) it is clear that the optimal stopping time is given by
\begin{eqnarray}
\rvt = \min\left\{1 \leq k \leq N - 1 | \Lambda_k \geq \tau_k \right\},
\end{eqnarray}
where the thresholds $\tau_k$ are characterized by (\ref{eqn:threshold}). Together with the optimal terminal decision rule $D$ obtained in (\ref{eqn:optimal-D}) we then prove Theorem \ref{thm:bayesian-structure}. $\Box$

In summary, the following \textbf{Algorithm BO-ODR} implements the Bayesian optimal ODR under maximum sample size $N$:
\vspace{0.1in}
\hrule
\vspace{0.1in}

\begin{center}
\textbf{Algorithm BO-ODR: Bayesian Optimal ODR under Maximum Sample Size $N$}
\end{center}

\noindent \textbf{Initial parameters:} Hypotheses $p_0, p_1$ and prior $\pi$, maximum sample size $N$, cost assignments $c_0, c_1, c$.

\noindent \textbf{Set:} A sequence of thresholds $\{\tau_k\}_{k = 1}^N$ computed via (\ref{eqn:threshold}) and $\tau_N = \frac{(1 - \pi) c_0}{\pi c_1}$.

\noindent \textbf{Algorithm:}

\hspace{0.1in}\textbf{initialize} $n = 1$;

\hspace{0.1in}\textbf{while} $n \leq N$

\hspace{0.3in}\textbf{do} compute $\Lambda_n$;

\hspace{0.3in}\textbf{if} $\Lambda_n \geq \tau_n$

\hspace{0.5in}\textbf{terminate} returning $\mathcal{H}_1$;

\hspace{0.3in}\textbf{else} $n = n + 1$;

\hspace{0.3in}\textbf{end if}

\hspace{0.1in}\textbf{end while}

\hspace{0.1in}\textbf{terminate} returning $\mathcal{H}_0$;

\vspace{0.1in}
\hrule
\vspace{0.2in}

We illustrate the Bayesian optimal ODR by the case study of testing the hypotheses
\begin{eqnarray}
\mathcal{H}_0: p_0 \sim \mathcal{N}(0, 1) \quad \mbox{versus}\quad \mathcal{H}_1: p_1 \sim \mathcal{N}(A, 1),
\end{eqnarray}
with $A > 0$.

\begin{figure}
\centering
\includegraphics[width=3.3in]{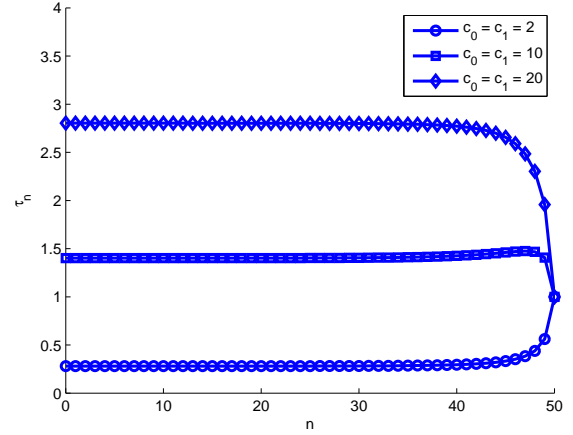}
\caption{Numerical results for the thresholds $\{\tau_n\}_{n = 1}^N$, for different $c_0$ and $c_1$.}
\label{fig:numerical-tau-cost}
\end{figure}

In the presented numerical examples (Figure \ref{fig:numerical-tau-cost}), we set $A = 1$, $\pi = 1/2$, $c = 1$, $N = 50$, and let $c_0 = c_1$ be $2$, $10$, and $20$ respectively. In the plots we display the thresholds $\{\tau_n\}_{n = 1}^N$, computed by \textbf{Algorithm BO-ODR}.
From the plots, we observe that the backward recursion quickly leads to stationary thresholds, within ten samples (returning from $n = N$). However, depending upon the values of tuning parameters (here the effective ones are $c_0 = c_1$), the evolution trend of the thresholds may differ considerably. In the plots, when $c_0 = c_1 = 2$ the sequence $\{\tau_n\}$ increases with $n$, when $c_0 = c_1 = 20$ it decreases with $n$, and when $c_0 = c_1 = 10$ there further exists a slight ``overshoot'' behavior. Intuitively, for small $c_0$ and $c_1$, the importance of reducing the expected stopping time under $\mathcal{H}_1$ outweighs that of decreasing the decision error probabilities, and hence it is reasonable to promote early stopping by using lower decision thresholds for early samples; on the contrary, for large $c_0$ and $c_1$, the priority is on decreasing the decision error probabilities, and hence it is reasonable to set relatively high decision thresholds for early samples, in order to avoid premature error-prone decisions.

\subsection{Case of Geometrically Distributed Maximum Sample Size}
\label{subsec:random}

In this subsection, we turn to the case in which the maximum sample size is no longer fixed, but is a geometrically distributed random variable $\rvn$ with a fixed finite mean $1/\epsilon$, i.e., $\mathrm{Pr}[\rvn = n] = (1 - \epsilon)^{n - 1} \epsilon$, $n = 1, 2, \ldots$. We assume that $\rvn$ is independent of $\rvx_1, \rvx_2, \ldots$. The realization of $\rvn$ is not revealed to the statistician until observing $\rvx_\rvn$: of course if the statistician has already made his opportunistic detection before observing $\rvx_\rvn$, there is no need to know about $\rvn$ any more; otherwise if the statistician has reached $\rvx_\rvn$ without a detection yet, then he is required to make his decision immediately with the $\rvn$ samples at hand, without observing any extra samples.

Due to its geometric distribution, $\rvn$ can be conveniently interpreted as the first time an i.i.d. sequence of Bernoulli trials (with success probability $\epsilon$) returns success. So alternatively $\rvn$ is a stopping time defined as follows:
\begin{eqnarray}
\rvn = \min\{n: \rvz_n = 1\},
\end{eqnarray}
where $\rvz_n$ is an i.i.d. sequence of Bernoulli random variables with $\mathrm{Pr}[\rvz_1 = 1] = \epsilon$ and $\mathrm{Pr}[\rvz_1 = 0] = 1 - \epsilon$. Therefore, for any stopping time $\rvt^\prime$ that is adapted to the filtration generated by $\rvx_1, \rvx_2, \ldots$, if we define
\begin{eqnarray}
\label{eqn:T-prime-2-T}
\rvt = \min\{\rvt^\prime, \rvn\},
\end{eqnarray}
then $\rvt$ is a stopping time adapted to the product filtration generated by $(\rvx_1, \rvz_1), (\rvx_2, \rvz_2), \ldots$. With a thus induced $\rvt$ and an arbitrary terminal decision rule $D$, we have the following definition of the ODR.
\begin{defn}
An ODR for a geometrically distributed maximum sample size $\rvn$ consists of a stopping time $\rvt$ given by (\ref{eqn:T-prime-2-T}) and a terminal decision rule $D: \mathcal{X}^\rvn \mapsto \{\mathcal{H}_0, \mathcal{H}_1\}$, such that, the statistician declares $\mathcal{H}_1$ if either $\{\rvt < \rvn\}$ or $\{\rvt = \rvn, D(\rvx^\rvn) = \mathcal{H}_1\}$ occurs, and declares $\mathcal{H}_0$ if $\{\rvt = \rvn, D(\rvx^\rvn) = \mathcal{H}_0\}$ occurs. Similar to that in Definition \ref{defn:ODR}, we may simply write $D(\rvx^\rvn)$ as $D$, when the context is unambiguous.
\end{defn}

Analogous to the problem framework in Section \ref{subsec:fixed-length}, we define the Bayesian cost as
\begin{eqnarray}
\label{eqn:bayesian-risk-random}
\mathcal{J} &=& (1 - \pi) c_0 P_\mathrm{FA} + \pi c_1 P_\mathrm{M} + c \mathbb{E}_1 [\rvt],
\end{eqnarray}
where $c_0, c_1, c > 0$ are cost assignments, and the problem then is to choose $\rvt$ and $D$ to minimize $\mathcal{J}$. Note that here the stopping time is not bounded since $\rvn$ can be arbitrarily large.

For the Bayesian cost, we have the following key fact.
\begin{prop}
\label{prop:Bayesian-reduction}
The Bayesian cost $\mathcal{J}$ in (\ref{eqn:bayesian-risk-random}) can be written in the following form:
\begin{eqnarray}
\label{eqn:Bayesian-reduction}
&&\mathcal{J} = \mathbb{E}_0\left[\sum_{n = 0}^{\rvt - 1} (1 - \epsilon)^n c \Lambda_n +\right.\nonumber\\
&&\left.(1 - \epsilon)^\rvt \left[(1 - \pi) c_0 + \frac{\epsilon}{1 - \epsilon} \min\{(1 - \pi) c_0, \pi c_1 \Lambda_\rvt \} \right]\right].\nonumber\\
\end{eqnarray}
\end{prop}

{\it Proof:} To proceed, consider the conditional Bayesian cost conditioned upon $\{\rvn = n\}$, $\mathcal{J}_n$; that is,
\begin{eqnarray}
\mathcal{J} = \sum_{n = 1}^\infty \mathrm{Pr}[\rvn = n] \mathcal{J}_n.
\end{eqnarray}
For evaluating the false alarm probability conditioned upon $\{\rvn = n\}$, we note that this conditional event is just $\{\rvt < n\} \cup \{\rvt = n, D = \mathcal{H}_1\}$. So we may write the conditional false alarm probability as
\begin{eqnarray}
P_{\mathrm{FA}, n} = \mathbb{E}_0[\mathbf{1}(\rvt < n)] + \mathbb{E}_0[\mathbf{1}(\rvt = n) \mathbf{1}(D = \mathcal{H}_1)].
\end{eqnarray}
Similarly, the conditional miss event is $\{\rvt = n, D = \mathcal{H}_0\}$, and we may write the conditional miss probability as
\begin{eqnarray}
P_{\mathrm{M}, n} &=& \mathbb{E}_1[\mathbf{1}(\rvt = n) \mathbf{1}(D = \mathcal{H}_0)]\nonumber\\
&=& \mathbb{E}_0[\Lambda_n \mathbf{1}(\rvt = n) \mathbf{1}(D = \mathcal{H}_0)],
\end{eqnarray}
since $\{\rvt = n, D = \mathcal{H}_0\}$ is $\mathcal{F}_n$-measurable.
So we have
\begin{eqnarray}
&&\mathcal{J}_n = \mathbb{E}_0\left[(1 - \pi) c_0 \mathbf{1}(\rvt < n) +\right.\nonumber\\
 &&(1 - \pi) c_0 \mathbf{1}(\rvt = n) \mathbf{1}(D = \mathcal{H}_1) + \nonumber\\
 &&\left. \pi c_1 \Lambda_n \mathbf{1}(\rvt = n) \mathbf{1}(D = \mathcal{H}_0) \right] + c \mathbb{E}_1[\rvt| \rvn = n].
\end{eqnarray}
Clearly for a given $\rvt$, for each $n$, the optimal $D$ that minimizes $\mathcal{J}_n$ is given by
\begin{eqnarray}
\label{eqn:optimal-D-n}
&&D = \mathcal{H}_1 \;\;\mbox{if}\; \Lambda_n \geq \frac{(1 - \pi) c_0}{\pi c_1};\nonumber\\
\mbox{and}\; &&D = \mathcal{H}_0 \;\;\mbox{otherwise}.
\end{eqnarray}
Since this solution does not depend on $n$, it is also the unconditional optimal terminal decision rule; that is,
\begin{eqnarray}
\label{eqn:optimal-D-random}
D = \mathcal{H}_1 \;\;\mbox{if}\; \Lambda_\rvn \geq \frac{(1 - \pi) c_0}{\pi c_1}; \;\mbox{and}\; D = \mathcal{H}_0 \;\;\mbox{otherwise}.
\end{eqnarray}

Now we can rewrite the Bayesian cost given the optimal terminal decision rule as
\begin{eqnarray}
\mathcal{J} = \sum_{n = 1}^\infty \mathrm{Pr}[\rvn = n] \mathbb{E}_0\left[(1 - \pi) c_0 \mathbf{1}(\rvt < n) +\right.\nonumber\\
 \left.\mathbf{1}(\rvt = n) \min\{(1 - \pi) c_0, \pi c_1 \Lambda_n \} \right] + c \mathbb{E}_1[\rvt].
\end{eqnarray}

Noting that
\begin{eqnarray}
&&\sum_{n = 1}^\infty \mathrm{Pr}[\rvn = n] \mathbb{E}_0\left[(1 - \pi) c_0 \mathbf{1}(\rvt < n)\right]\nonumber\\
&=& (1 - \pi) c_0 \mathbb{E}_0\left[\sum_{n = 1}^\infty \mathrm{Pr}[\rvn = n] \mathbf{1}(\rvt < n)\right]\nonumber\\
&=& (1 - \pi) c_0 \mathbb{E}_0\left[\sum_{n = \rvt + 1}^\infty \mathrm{Pr}[\rvn = n] \right]\nonumber\\
&=& (1 - \pi) c_0 \mathbb{E}_0\left[\sum_{n = \rvt + 1}^\infty (1 - \epsilon)^{n - 1} \epsilon \right]\nonumber\\
&=& (1 - \pi) c_0 \mathbb{E}_0\left[(1 - \epsilon)^\rvt \right],
\end{eqnarray}
we have
\begin{eqnarray}
&&\mathcal{J} = \mathbb{E}_0\left[(1 - \pi) c_0 (1 - \epsilon)^\rvt +\right.\nonumber\\
&& \left.(1 - \epsilon)^{\rvt - 1} \epsilon \min\{(1 - \pi) c_0, \pi c_1 \Lambda_\rvt \}\right] + c \mathbb{E}_1[\rvt].
\end{eqnarray}

Next we evaluate $\mathbb{E}_1[\rvt]$, as follows:
\begin{eqnarray}
\mathbb{E}_1[\rvt] &\stackrel{(a)}{=}& \mathbb{E}_1[\min\{\rvt, \rvn\}]\nonumber\\
&\stackrel{(b)}{=}& \sum_{n = 1}^\infty (1 - \epsilon)^{n - 1} \epsilon \mathbb{E}_1[\min\{\rvt, n\}]\nonumber\\
&\stackrel{(c)}{=}& \mathbb{E}_1\left[\sum_{n = 1}^\rvt n (1 - \epsilon)^{n - 1} \epsilon + \sum_{n = \rvt + 1}^\infty \rvt (1 - \epsilon)^{n - 1} \epsilon\right]\nonumber\\
&\stackrel{(d)}{=}& \mathbb{E}_1\left[\frac{1 - (1 - \epsilon)^\rvt}{\epsilon}\right] = \mathbb{E}_1\left[\sum_{n = 0}^{\rvt - 1} (1 - \epsilon)^n \right]\nonumber\\
&\stackrel{(e)}{=}& \mathbb{E}_0\left[\sum_{n = 0}^{\rvt - 1} (1 - \epsilon)^n \Lambda_n\right],
\end{eqnarray}
where, (a) is due to the fact that $\rvt$ is upper bounded by $\rvn$, (b) is the total expectation expansion, (c) is due to the fact that the expectation of $\rvt$ is bounded (by the expectation of $\rvn$), (d) is obtained via algebraic manipulations, and (e) is due to the fact that the event $\{n \leq \rvt - 1\}$ is $\mathcal{F}_n$-measurable.

So, back to the Bayesian cost, we have reached (\ref{eqn:Bayesian-reduction}) and thus proved Proposition \ref{prop:Bayesian-reduction}. $\Box$

An inspection of (\ref{eqn:Bayesian-reduction}) reveals that it is exactly the form that has been treated in \cite[2.14]{shiryaev78:book}, considering both an instantaneous reward at the stopping time and accumulated sampling costs, with everything discounted by an exponential factor $(1 - \epsilon)^n$ at time $n$.

Let us define for $\lambda \geq 0$
\begin{eqnarray}
g(\lambda) = (1 - \pi) c_0 + \frac{\epsilon}{1 - \epsilon} \min\{(1 - \pi) c_0, \pi c_1 \lambda\},
\end{eqnarray}
and $c(\lambda) = c\lambda$. First, the relevant regularity conditions \cite[(2.168)]{shiryaev78:book} hold, namely that $|g(\lambda)|$ is finitely bounded, and that $\mathbb{E}_0[c(\Lambda_n)]$ is finite for every $n$. So, as a consequence of \cite[Thm. 23]{shiryaev78:book}, we have the following result.
\begin{thm}
The Bayesian optimal stopping time is given by
\begin{eqnarray}
\label{eqn:opt-stop-random}
\rvt = \min\{n \geq 1: V(\Lambda_n) = g(\Lambda_n)\},
\end{eqnarray}
where $V(\cdot)$ is the solution of
\begin{eqnarray}
\label{eqn:stationary-state}
V(\lambda) = \min\{g(\lambda), (1 - \epsilon)\mathbb{E}_0[V(\lambda \rvl)] + c(\lambda)\},
\end{eqnarray}
with $\rvl = p_1(\rvx)/p_0(\rvx)$, $\rvx$ following $p_0$. Furthermore, $V(\cdot)$ may be computed as $V(\lambda) = \lim_{n \rightarrow \infty} \mathcal{Q}^n g(\lambda)$, with the operator $\mathcal{Q}$ defined by
\begin{eqnarray}
\mathcal{Q} f(\lambda) = \min\{f(\lambda), (1 - \epsilon) \mathbb{E}_0[f(\lambda \rvl)] + c(\lambda)\}.
\end{eqnarray}
\end{thm}

The Bayesian optimal stopping time (\ref{eqn:opt-stop-random}) leads to a likelihood ratio threshold test, as given by the following result.

\begin{cor}
\label{cor:geo-stop-lrt}
Define the ``running'' threshold $\tau_r$ as the value of $\lambda$ at the intersection of $g(\lambda)$ and $(1 - \epsilon) \mathbb{E}_0[V(\lambda\rvl)] + c(\lambda)$, which always exists and is unique, and the ``terminal'' threshold $\tau_t = (1 - \pi)c_0/(\pi c_1)$. The Bayesian optimal stopping rule is described by \textbf{Algorithm BO-ODR-Geo}.
\end{cor}

{\it Proof:} It suffices to prove that $\tau_r$ always exists and is unique; that is, $g(\lambda)$ and $(1 - \epsilon) \mathbb{E}_0[V(\lambda\rvl)] + c(\lambda)$ intersect only once. By induction, it follows that $V(\lambda)$ is a monotonically non-decreasing and concave continuous function of $\lambda > 0$, and that $\lim_{\lambda \rightarrow 0^+} V(\lambda) = 0$. Therefore, $(1 - \epsilon) \mathbb{E}_0[V(\lambda\rvl)] + c(\lambda)$ grows without bound as $\lambda \rightarrow \infty$, and hence it must intersect at least once at $g(\lambda)$ over $\lambda > 0$. To prove that the intersection is unique, we note that from (\ref{eqn:stationary-state})
\begin{eqnarray}
V(\lambda) &=& \min\{g(\lambda), (1 - \epsilon)\mathbb{E}_0[V(\lambda \rvl)] + c(\lambda)\}\nonumber\\
&\leq& (1 - \epsilon)\mathbb{E}_0[V(\lambda \rvl)] + c(\lambda) \nonumber\\
&\stackrel{(a)}{\leq}& (1 - \epsilon) V(\lambda \mathbb{E}_0[\rvl]) + c(\lambda)\nonumber\\
&\stackrel{(b)}{=}& (1 - \epsilon) V(\lambda) + c(\lambda),
\end{eqnarray}
that is,
\begin{eqnarray}
V(\lambda) \stackrel{(c)}{\leq} \frac{c\lambda}{\epsilon},
\end{eqnarray}
wherein, (a) is from the concavity of $V(\cdot)$, (b) is from the fact that $\mathbb{E}_0[\rvl] = 1$, and (c) is from the fact that $c(\lambda) = c \lambda$.

We then consider two cases.

Case 1: $\epsilon > c/(c + \pi c_1)$. In this case, it is impossible for $(1 - \epsilon) \mathbb{E}_0[V(\lambda\rvl)] + c(\lambda)$ to intersect $g(\lambda)$ for any $\lambda < \tau_t = (1 - \pi)c_0/(\pi c_1)$, or, it is only possible for $(1 - \epsilon) \mathbb{E}_0[V(\lambda\rvl)] + c(\lambda)$ to intersect $g(\lambda)$ at some $\lambda \geq \tau_t$, for which $g(\lambda) = (1 - \pi)c_0/(1 - \epsilon)$ is a horizontal line, and thus the intersection is unique. To see this, note that in the case of $\epsilon > c/(c + \pi c_1)$, we have
\begin{eqnarray}
V(\tau_t) \leq \frac{c \tau_t}{\epsilon} = \frac{c (1 - \pi) c_0}{\epsilon \pi c_1} < \frac{(1 - \pi) c_0 (c + \pi c_1)}{\pi c_1},
\end{eqnarray}
and
\begin{eqnarray}
g(\tau_t) = \frac{(1 - \pi)c_0}{1 - \epsilon} > \frac{(1 - \pi) c_0 (c + \pi c_1)}{\pi c_1}.
\end{eqnarray}

Case 2: $\epsilon \leq c/(c + \pi c_1)$. In this case, it is possible for $(1 - \epsilon) \mathbb{E}_0[V(\lambda\rvl)] + c(\lambda)$ to intersect $g(\lambda)$ for some $\lambda_1 < \tau_t$. But if this happens, then it is impossible for these two curves to intersect for any other $\lambda > \lambda_1$ and thus the intersection is also unique. To see this, note that the slope of $(1 - \epsilon) \mathbb{E}_0[V(\lambda\rvl)] + c(\lambda)$ is always lower bounded by $c$, while the slope of $g(\lambda)$ for $\lambda < \tau_t$ is $\epsilon \pi c_1/(1 - \epsilon)$, which is no greater than $c$ in the case of $\epsilon \leq c/(c + \pi c_1)$.

Summarizing Cases 1 and 2, we conclude the proof of Corollary \ref{cor:geo-stop-lrt}.
$\Box$

\vspace{0.1in}
\hrule
\vspace{0.1in}

\begin{center}
\textbf{Algorithm BO-ODR-Geo: Bayesian Optimal ODR under Geometrically Distributed Maximum Sample Size}
\end{center}

\noindent \textbf{Initial parameters:} Hypotheses $p_0, p_1$ and prior $\pi$, mean sample size $1/\epsilon$, cost assignments $c_0, c_1, c$.

\noindent \textbf{Set:} The ``running'' threshold $\tau_r$ and the ``terminal'' threshold $\tau_t$, according to Corollary \ref{cor:geo-stop-lrt}.

\noindent \textbf{Algorithm:}

\hspace{0.1in}\textbf{initialize} $n = 1$;

\hspace{0.1in}\textbf{while} $\rvn$ has not been revealed

\hspace{0.3in}\textbf{do} compute $\Lambda_n$;

\hspace{0.3in}\textbf{if} $\Lambda_n \geq \tau_r$

\hspace{0.5in}\textbf{terminate} returning $\mathcal{H}_1$;

\hspace{0.3in}\textbf{else} $n = n + 1$;

\hspace{0.3in}\textbf{end if}

\hspace{0.1in}\textbf{end while}

\hspace{0.1in}\textbf{if} $\Lambda_\rvn \geq \tau_t$

\hspace{0.3in}\textbf{terminate} returning $\mathcal{H}_1$;

\hspace{0.1in}\textbf{else}

\hspace{0.3in}\textbf{terminate} returning $\mathcal{H}_0$;

\hspace{0.1in}\textbf{end if}

\vspace{0.1in}
\hrule
\vspace{0.2in}

For the optimal ODR, an interesting property is that it is a two-threshold scheme: the ``running'' threshold $\tau_r$, which is determined by solving the stationary state equation (\ref{eqn:stationary-state}), is used to compare with the likelihood ratio sequence before $\rvn$, i.e., when future samples are still available; and the ``terminal'' threshold $\tau_t$, which is simply the ratio between the priors scaled by costs, is used only at the end, i.e., when the statistician is informed that the final sample has been reached and a decision is required immediately. Such a two-threshold scheme is very different from the conventional one-sided and two-sided SPRTs, in which the thresholds are fixed constants throughout.

\begin{figure}
\centering
\includegraphics[width=3.3in]{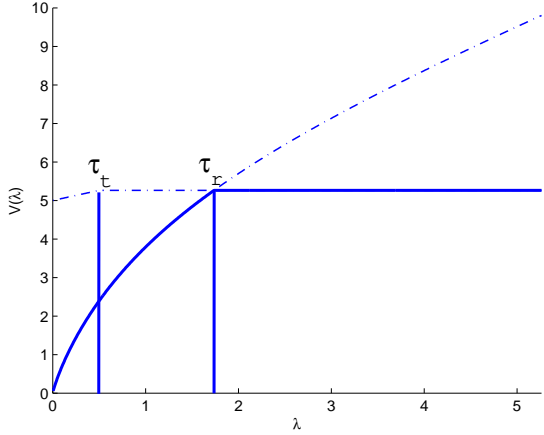}
\caption{Illustration of typical situation for $\tau_r, \tau_t$, with $c_0 = 10, c_1 = 20$.}
\label{fig:geo_stop_illustration1}
\end{figure}

\begin{figure}
\centering
\includegraphics[width=3.3in]{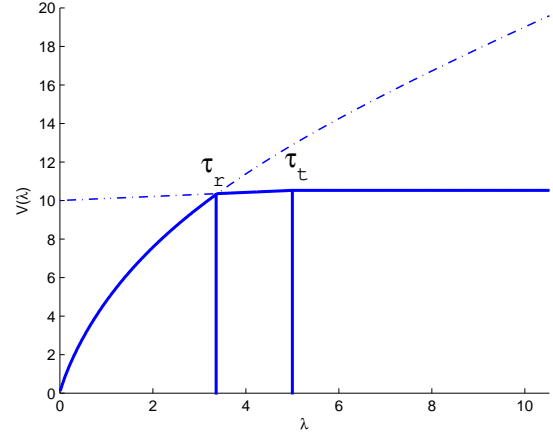}
\caption{Illustration of typical situation for $\tau_r, \tau_t$, with $c_0 = 20, c_1 = 4$.}
\label{fig:geo_stop_illustration2}
\end{figure}

\begin{figure}
\centering
\includegraphics[width=3.3in]{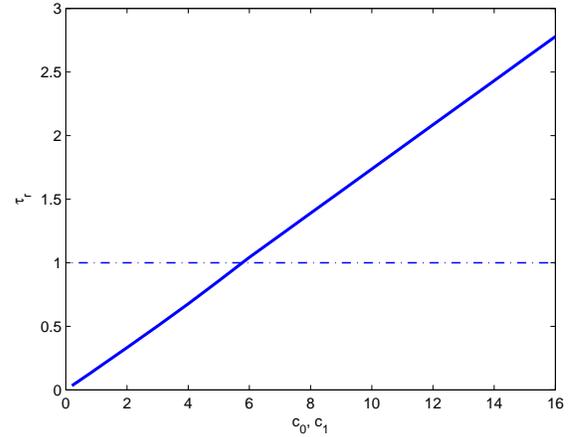}
\caption{The running threshold $\tau_r$ versus $c_0$ and $c_1$ (herein $c_0 = c_1$). The dash-dot line indicates the terminal threshold $\tau_t$.}
\label{fig:geo_stop_tauR}
\end{figure}

We use the same case study as that considered in Section \ref{subsec:fixed-length} to illustrate the numerical behavior of the optimal ODR under geometrically distributed maximum sample size. Again we set $A = 1$, $\pi = 1/2$, and $c = 1$. For the geometric distribution of $\rvn$, we set $\epsilon = 0.05$, so that the mean maximum sample size is $20$. Note that $g(\lambda)$ is a piecewise linear function of $\lambda$ with one switching point exactly at $\lambda = \tau_t$; so depending on at which segment the curve $(1 - \epsilon) \mathbb{E}_0[V(\lambda \rvl)] + c(\lambda)$ intersects $g(\lambda)$, there are two possible situations, as illustrated in Figures \ref{fig:geo_stop_illustration1} and \ref{fig:geo_stop_illustration2}, respectively. In the former, $\tau_r \geq \tau_t$, and in the latter, $\tau_r < \tau_t$.
In Figure \ref{fig:geo_stop_tauR} we plot the trend of $\tau_r$ as $c_0 = c_1$ increases from $0.2$ to $16$. We observe that $\tau_r$ increases with $c_0$ and $c_1$, crossing the level of $\tau_t$. Interestingly, the growth trend of $\tau_r$ is virtually linear with $c_0$ and $c_1$.

\section{Asymptotic Analysis}
\label{sec:tradeoff}

In the previous section, we have focused on the Bayesian optimal stopping rules, which yield ODRs that minimize corresponding Bayesian costs. In this section, we turn to the asymptotic regime, letting the maximum sample size $N$ grow without bound. The performance metrics here are, instead of the Bayesian cost, the exponential decay rates of the (false alarm and miss) error probabilities and the expected stopping time under $\mathcal{H}_1$.

In general, for a sequence of ODRs indexed by the maximum sample size $N = 1, 2, \ldots$, we have an asymptotic tradeoff among three performance metrics: the exponential decay rate of the false alarm probability, the exponential decay rate of the miss probability, and the expected stopping time (normalized by $N$) under $\mathcal{H}_1$. Mathematically a performance tuple $(\Delta_\mathrm{FA}, \Delta_\mathrm{M}, \eta)$ is achievable if there exists a sequence of ODRs indexed by $N$, such that
\begin{eqnarray}
&&\liminf_{N \rightarrow \infty} \frac{-\log P_\mathrm{FA}}{N} \geq \Delta_\mathrm{FA},\\
&&\liminf_{N \rightarrow \infty} \frac{-\log P_\mathrm{M}}{N} \geq \Delta_\mathrm{M},\\
&&\limsup_{N \rightarrow \infty} \frac{T}{N} \leq \eta,
\end{eqnarray}
where $T = \mathbb{E}_1[\rvt]$ is the expected stopping time under $\mathcal{H}_1$.

Furthermore, we may call the closure of the union of achievable tuples under all possible ODRs the ODR performance region, which should depend solely upon $(p_0, p_1)$. We denote the ODR performance region by $\mathcal{R}(p_0, p_1)$, which is a subset of $[0, \infty) \times [0, \infty) \times [0, 1] \subset \mathbb{R}^3$.

The following theorem is the main result of our asymptotic analysis, which fully characterizes $\mathcal{R}(p_0, p_1)$.

\begin{thm}
\label{thm:odr-region}
The ODR performance region $\mathcal{R}(p_0, p_1)$ is given as follows: for each $0 \leq \eta \leq 1$,
\begin{eqnarray}
\Delta_\mathrm{FA} &\leq& \min\left\{\eta d_1, \sup_{\alpha > 0} \left\{\alpha \left[d_1 - \nu (d_0 + d_1)\right] -\right.\right.\nonumber\\
&& \left.\left.\log \mathbb{E}_0\left[e^{\alpha \log p_1(\rvx)/p_0(\rvx)}\right]
\right\}\right\},\nonumber\\
\Delta_\mathrm{M} &\leq& \sup_{\alpha < 0} \left\{\alpha \left[d_1 - \nu (d_0 + d_1)\right] -\right.\nonumber\\
&& \left.\log \mathbb{E}_1\left[e^{\alpha \log p_1(\rvx)/p_0(\rvx)}\right]
\right\},
\end{eqnarray}
for $0 \leq \nu \leq 1$, where $d_0 = D(p_0\|p_1)$ and $d_1 = D(p_1\|p_0)$.
\end{thm}

Theorem \ref{thm:odr-region} is proved in two parts. The achievability part is established by constructing a specific form of ODRs that asymptotically achieve the performance tuple as described in Theorem \ref{thm:odr-region}. The converse part is established by an argument of contradiction, in which a key idea is information-theoretic, basically asserting that, if the ODR performance region $\mathcal{R}(p_0, p_1)$ can be outperformed, then one can achieve a rate per unit cost higher than the capacity per unit cost \cite{verdu90:it} for a certain stationary memoryless channel, an impossible task even with feedback and variable-length coding \cite{csiszar73:pcit}. The detailed steps of the proof are given in Sections \ref{subsec:stein} through \ref{subsec:completion}. In the next two subsections we provide some illustration and discussion of Theorem \ref{thm:odr-region}.

\subsection{Case Study: Gaussian Distributions with and without a Drift}

To illustrate the ODR performance region in Theorem \ref{thm:odr-region}, we present a case study for the following hypotheses:
\begin{eqnarray}
\label{eqn:hypotheses-brownian-case-study}
\mathcal{H}_0: p_0 \sim \mathcal{N}(0, 1) \quad \mbox{versus}\quad \mathcal{H}_1: p_1 \sim \mathcal{N}(A, 1),
\end{eqnarray}
with $A > 0$. In this case we have $D(p_0\|p_1) = D(p_1\|p_0) = A^2/2$.

\begin{figure}
\centering
\includegraphics[width=3.3in]{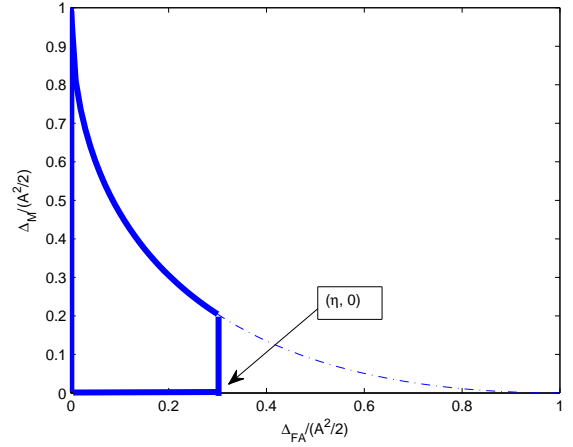}
\caption{An illustration of (\ref{eqn:tuple-cut}).}
\label{fig:tuple-cut}
\end{figure}

Then, applying Theorem \ref{thm:odr-region}, we can obtain the (normalized) region $\left(\frac{\Delta_\mathrm{FA}}{A^2/2}, \frac{\Delta_\mathrm{M}}{A^2/2}\right)$ for every fixed $0 \leq \eta \leq 1$, as
\begin{eqnarray}
\label{eqn:tuple-cut}
\left\{\left(\frac{\Delta_\mathrm{FA}}{A^2/2}, \frac{\Delta_\mathrm{M}}{A^2/2}\right) = (x, y):\right.\nonumber\\
 \left.\sqrt{x} + \sqrt{y} \leq 1, 0 \leq x \leq \eta, y \geq 0  \right\},
\end{eqnarray}
illustrated in Figure \ref{fig:tuple-cut}.

\begin{figure}
\centering
\includegraphics[width=3.3in]{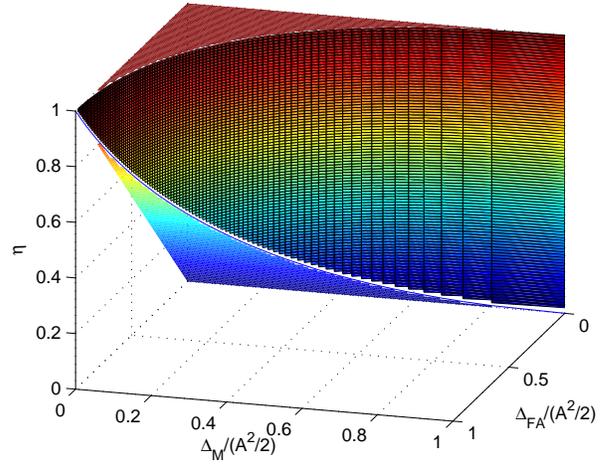}
\caption{An illustration of $\mathcal{R}(p_0, p_1)$ in (\ref{cor:tuple-region}).}
\label{fig:tuple-region}
\end{figure}

The complete characterization of $\mathcal{R}(p_0, p_1)$ is given by the following corollary and illustrated in Figure \ref{fig:tuple-region}.
\begin{cor}
For the hypotheses (\ref{eqn:hypotheses-brownian-case-study}), the ODR performance region $\mathcal{R}(p_0, p_1)$ is
\begin{eqnarray}
\label{cor:tuple-region}
\left\{\left(\frac{\Delta_\mathrm{FA}}{A^2/2}, \frac{\Delta_\mathrm{M}}{A^2/2}, \eta\right) = (x, y, z):\right.\nonumber\\
\left. \sqrt{x} + \sqrt{y} \leq 1, 0 \leq x \leq z, y \geq 0, 0 \leq z \leq 1  \right\}.
\end{eqnarray}
\end{cor}

\subsection{Stein-Chernoff Lemma Revisited}

In this subsection, we focus on an extremal case of Theorem \ref{thm:odr-region}, in which the false alarm probability is fixed without decreasing toward zero exponentially, or, has an exponent of zero. For this case, Theorem \ref{thm:odr-region} specializes into the following corollary.

\begin{cor}
\label{cor:odr-exponent}
For an arbitrary fixed target false alarm probability $P^\ast_\mathrm{FA} > 0$, among all sequences of ODRs such that the normalized expected stopping time under $\mathcal{H}_1$, $T/N$, satisfies $\lim_{N \rightarrow \infty} {T}/{N} = 0$, the maximum achievable error exponent of $P_\mathrm{M}$ is
\begin{eqnarray}
\lim_{N \rightarrow \infty} \frac{-\log P_\mathrm{M}}{N} = D(p_0\|p_1).
\end{eqnarray}
\end{cor}

A similar situation has been treated in \cite{zhang13:it}, wherein the considered form of ODR is restricted to be a truncated one-sided SPRT, that is,
\begin{eqnarray}
\label{eqn:fodr-representation}
\hat{\mathcal{H}} = \mathbf{1}\left(\bigcup_{k = 1}^N \{\Lambda_k \geq B\}\right),
\end{eqnarray}
where $\mathbf{1}(\cdot)$ is the indicator function. It was shown therein that the above form of ODR behaves asymptotically according to the following theorem.

\begin{thm}
\label{thm:odr-fixed}
(\cite[Thm. 1, Thm. 2]{zhang13:it}) For the truncated one-sided SPRT ODR of the form (\ref{eqn:fodr-representation}) that attains an arbitrary fixed target false alarm probability $0 < P^\ast_\mathrm{FA} \leq P_0[p_1(\rvx) \geq p_0(\rvx)]$, the miss probability scales toward zero as $N$ grows without bound as
\begin{eqnarray}
\lim_{N \rightarrow \infty} \frac{-\log P_\mathrm{M}}{N} = C(p_0, p_1),
\end{eqnarray}
where $C(p_0, p_1)$ is the Chernoff information of $(p_0, p_1)$ (see \cite{chernoff52:ams} and \cite[Ch. 11.9]{cover06:book})
\begin{eqnarray}
C(p_0, p_1) = -\inf_{\alpha \in (0, 1)} \log \left(\int_{\mathcal{X}} p_0^\alpha(x) p_1^{1 - \alpha}(x) dx\right),
\end{eqnarray}
and the normalized expected stopping time under $\mathcal{H}_1$, $T/N$, satisfies
\begin{eqnarray}
\lim_{N \rightarrow \infty} \frac{T}{N} = 0.
\end{eqnarray}
\end{thm}

Comparing Theorem \ref{thm:odr-fixed} and Corollary \ref{cor:odr-exponent}, we can conclude that, among all sequences of ODRs such that $\lim_{N \rightarrow \infty} T/N = 0$, there exist ODRs that achieve a larger error exponent of $P_\mathrm{M}$ than that achieved by the truncated one-sided SPRT ODR in \cite{zhang13:it}.
\footnote{It should be noted that the comparison is based on the footing of $\lim_{N \rightarrow \infty} T/N = 0$. It is possible that for finer asymptotic behaviors, the comparison becomes more delicate; for example, one may ask whether the conclusion still holds if one focuses on sequences of ODRs such that $T = O(N^\alpha)$ for some $0 < \alpha < 1$, and the result is unknown.}
The error exponent achieved in Corollary \ref{cor:odr-exponent}, $D(p_0\|p_1)$, is exactly that achieved by the optimal FSS decision rule as indicated in the Stein-Chernoff Lemma, but here the corresponding ODR is not FSS, only requiring asymptotically diminishing sampling cost under $\mathcal{H}_1$. So in other words, the FSS sampling cost is not fundamental in achieving the Stein-Chernoff Lemma, which appears to be a new and somewhat surprising finding.

\subsection{Proof of Corollary \ref{cor:odr-exponent}}
\label{subsec:stein}

Let us represent an ODR in a general form beyond that in (\ref{eqn:fodr-representation}) as
\begin{eqnarray}
\hat{\mathcal{H}}(\underline{f}, \underline{B}) = \mathbf{1}\left(\bigcup_{k = 1}^N \left\{f_k(\rvx_1, \ldots, \rvx_k) \geq B_k\right\}\right),
\end{eqnarray}
where $\underline{f}$ is a sequence of processing functions, and $\underline{B}$ is a sequence of thresholds. Note that $\hat{\mathcal{H}}(\underline{f}, \underline{B})$ also includes as a special case the FSS Neyman-Pearson decision rules, which have $\underline{f}$ as likelihood ratios, and $\underline{B}$ as $B_k = \infty$ for all $k$ except for $k = N$.

There are several ways of constructing an ODR that achieves the asymptotic performance in Corollary \ref{cor:odr-exponent}. Here we give a proof based on a simple idea of two-stage ODRs, which may not be the most sensible choice for finite $N$ in practice but is sufficient for proving the asymptotic result and is quite convenient to analyze. That is, we restrict the sequence of processing functions, $\underline{f}$, to be likelihood ratios, and among the elements of the threshold sequence $\underline{B}$, we only let two of them be finite, i.e.,
\begin{eqnarray}
&&B_M = e^{\tau_M}, \quad B_N = e^{\tau_N},\nonumber\\
\mbox{and}\;&& B_k = \infty \quad \mbox{for}\; k \neq M, N,
\end{eqnarray}
where $M < N$ corresponds to an early stage at which an opportunistic decision may be made, and $e^{\tau_{M, N}}$ are the thresholds for the two stages. We identify such two-stage ODRs with the designation $2$-ODR, and sometimes represent them with the notation $\hat{\mathcal{H}}_2(M, \tau_M, \tau_N)$. So in words, for $2$-ODRs there is only one opportunity (upon observing the first $M$ samples) to stop early.

In the proof of Corollary \ref{cor:odr-exponent}, we fix $M/N = \epsilon > 0$, and let $\tau_M = -M \left[D(p_0\|p_1) - \delta\right]$ and $\tau_N = -N \left[D(p_0\|p_1) - \delta\right]$ for some small $\delta > 0$. The miss probability is thus bounded as
\begin{eqnarray}
P_\mathrm{M} &=& P_1\left[\Lambda_M < e^{\tau_M}, \Lambda_N < e^{\tau_N}\right]\nonumber\\
&\leq& P_1\left[\log \Lambda_N < \tau_N\right]\nonumber\\
&\leq& \exp\left\{\inf_{0 \leq \alpha \leq 1} \left[-\alpha N (D(p_0\|p_1) - \delta) +\right.\right.\nonumber\\
 &&\quad\quad\quad \left.\left.\log \mathbb{E}_1\left[e^{-\alpha \log \Lambda_N}\right]\right]\right\}\nonumber\\
&=& \exp\left\{-N \sup_{0 \leq \alpha \leq 1} \left[\alpha D(p_0\|p_1) - \alpha \delta -\right.\right.\nonumber\\
 &&\quad\quad\quad \left.\left.\log \mathbb{E}_1\left[e^{-\alpha \log p_1(\rvx)/p_0(\rvx)}\right] \right]\right\}\nonumber\\
&\leq& \exp\left\{-N \left[D(p_0\|p_1) - \delta\right]\right\},
\end{eqnarray}
by letting $\alpha = 1$.

The false alarm probability is bounded as
\begin{eqnarray}
\label{eqn:fa-bound}
P_\mathrm{FA} &=& P_0\left[\Lambda_M \geq e^{\tau_M} \;\mbox{or}\; \Lambda_N \geq e^{\tau_N}\right]\nonumber\\
&\leq& P_0\left[\log \Lambda_M \geq \tau_M\right] + P_0\left[\log \Lambda_N \geq \tau_N\right]
\end{eqnarray}
due to the union bound. From the weak law of large numbers, both probabilities in (\ref{eqn:fa-bound}) approach zero for any fixed $\delta > 0$, as $N$ grows without bound. Thus $P_\mathrm{FA}$ can be ensured to be arbitrarily small as $N$ grows without bound.

Regarding the expected stopping time under $\mathcal{H}_1$, we have
\begin{eqnarray}
T &=& M \cdot P_1\left[\Lambda_M \geq e^{\tau_M}\right] + N \cdot P_1\left[\Lambda_M < e^{\tau_M}\right]\nonumber\\
&\leq& M + N \cdot \exp\left\{-M \left[D(p_0\|p_1) - \delta\right]\right\}.
\end{eqnarray}
So it follows that
\begin{eqnarray}
\frac{T}{N} \leq \frac{M}{N} + \exp\left\{-M \left[D(p_0\|p_1) - \delta\right]\right\},
\end{eqnarray}
which converges as $N \rightarrow \infty$ to $\epsilon = M/N$. So the proof of Corollary \ref{cor:odr-exponent} is completed by letting $\delta \rightarrow 0$ and $\epsilon \rightarrow 0$. $\Box$

\subsection{The $\left(\Delta_\mathrm{M} = 0, \Delta_\mathrm{FA} = \eta D(p_1\|p_0)\right)$ Corner Point}
\label{subsec:corner}

According to Theorem \ref{thm:odr-region}, a boundary of $\mathcal{R}(p_0, p_1)$ is given by $\Delta_\mathrm{FA} = \eta D(p_1\|p_0)$ for every $0 \leq \eta \leq 1$. This extremal case corner point of $\left(\Delta_\mathrm{M} = 0, \Delta_\mathrm{FA} = \eta D(p_1\|p_0)\right)$ specializes into the following corollary.

\begin{cor}
\label{cor:odr-exponent-falsealarm}
For an arbitrary fixed target miss probability $P^\ast_\mathrm{M} > 0$, there exists a sequence of ODRs such that when the normalized expected stopping time under $\mathcal{H}_1$, $T/N$, satisfies
\begin{eqnarray}
\label{eqn:stopping-time-falsealarm}
\lim_{N \rightarrow \infty} \frac{T}{N} = \eta \leq 1,
\end{eqnarray}
the false alarm probability satisfies
\begin{eqnarray}
\label{eqn:odr-exponent-falsealarm}
\lim_{N \rightarrow \infty} \frac{-\log P_\mathrm{FA}}{N} = \eta D(p_1\|p_0).
\end{eqnarray}
Furthermore, no ODR may achieve a larger exponent for $P_\mathrm{FA}$ under the constraint of (\ref{eqn:stopping-time-falsealarm}) on $T$.
\end{cor}

{\it Proof of the Achievability Part:} To prove the existence of ODRs that achieve (\ref{eqn:stopping-time-falsealarm}) and (\ref{eqn:odr-exponent-falsealarm}), consider $2$-ODRs $\left\{\hat{\mathcal{H}}_2(M, \tau_M, \tau_N)\right\}$, in which we set
\begin{eqnarray}
&&M = \eta N, \quad \tau_M = M \left[D(p_1\|p_0) - \delta\right],\nonumber\\
&& \tau_N = N \left[D(p_1\|p_0) - \delta\right],
\end{eqnarray}
for some small $\delta > 0$. The miss probability $P_\mathrm{M}$ satisfies
\begin{eqnarray}
P_\mathrm{M} &=& P_1\left[\log \Lambda_M < \tau_M, \log \Lambda_N < \tau_N \right]\nonumber\\
&\leq& P_1\left[\log \Lambda_N < \tau_N \right],
\end{eqnarray}
which can be ensured to be arbitrarily small as $N$ grows without bound, for any fixed $\delta > 0$. For $T$, we have
\begin{eqnarray}
\frac{T}{N} &=& \frac{M}{N} \cdot P_1\left[\log \Lambda_M \geq \tau_M\right] + P_1\left[\log \Lambda_M < \tau_M \right]\nonumber\\
&\leq& \eta + P_1\left[\log \Lambda_M < \tau_M \right],
\end{eqnarray}
which converges to $\eta$ as $N$ grows without bound, for any fixed $\delta > 0$.

For the false alarm probability $P_\mathrm{FA}$, we have
\begin{eqnarray}
P_\mathrm{FA} &=& P_0\left[\log \Lambda_M \geq \tau_M \;\mathrm{or}\; \log \Lambda_N \geq \tau_N\right]\nonumber\\
&\leq& P_0\left[\log \Lambda_M \geq \tau_M\right] + P_0\left[\log \Lambda_N \geq \tau_N\right]\nonumber\\
&\leq& \exp\left\{-M \sup_{0 \leq \alpha \leq 1} \left[\alpha D(p_1\|p_0) - \alpha \delta - \right.\right.\nonumber\\
&&\left.\left.\log \mathbb{E}_0\left[e^{\alpha \log p_1(\rvx)/p_0(\rvx)}\right]\right]\right\} + \nonumber\\
&&\quad \exp\left\{-N \sup_{0 \leq \alpha \leq 1} \left[\alpha D(p_1\|p_0) - \alpha \delta - \right.\right.\nonumber\\
&&\left.\left.\log \mathbb{E}_0\left[e^{\alpha \log p_1(\rvx)/p_0(\rvx)}\right]\right]\right\}\nonumber\\
&\leq& \exp\left\{-\eta N \left[D(p_1\|p_0) - \delta\right]\right\} +\nonumber\\
&& \quad\quad \exp\left\{-N \left[D(p_1\|p_0) - \delta\right]\right\},
\end{eqnarray}
by setting $\alpha = 1$ in both exponents. Herein, the first term dominates the exponential decay behavior as $N$ grows without bound. So the achievability part of Corollary \ref{cor:odr-exponent-falsealarm} is established by letting $\delta \rightarrow 0$.

{\it Proof of the Converse Part:} To prove that there are no ODRs that outperform the asymptotic performance specified in (\ref{eqn:stopping-time-falsealarm}) and (\ref{eqn:odr-exponent-falsealarm}), we use the argument of contradiction, which borrows ideas from the information-theoretic analysis of the channel capacity per unit cost \cite{verdu90:it}.

Here we briefly describe the channel capacity per unit cost problem, simplified for our problem setup. Consider a stationary memoryless channel. Let the channel input alphabet $\mathcal{S}$ consist of two letters, $s_0$ and $s_1$, whose corresponding conditional output distributions are $p_0(x)$ and $p_1(x)$, respectively. We assume that the cost of using $s_0$ as channel input is zero, and that of using $s_1$ is one. As established in \cite[Thm. 3]{verdu90:it}, the channel capacity per unit cost of this channel is given by $\mathbf{C} = D(p_1\|p_0)$. Furthermore, from \cite{csiszar73:pcit}, $\mathbf{C}$ remains the channel capacity per unit cost even in the presence of feedback and variable-length coding.

Then, for the channel model above, consider the following encoding/decoding scheme. Denote the size of the message set by $M$, in which a message is selected uniformly at random for transmission, and introduce a parameter $N$. A ``root'' codebook is constructed as a collection of $M$ different $M \times N$ matrices, wherein the $m$th message corresponds to an $M \times N$ codeword matrix whose $m$th-row elements are all $s_1$, and the remaining elements in the matrix are all $s_0$.

When no feedback is available, the root codebook is the actual codebook used for transmission \cite[pp. 1023-1024]{verdu90:it}. That is, once a message $m$ is selected, the encoder transmits the corresponding matrix, row by row. The decoder conducts a binary hypothesis test for each received row; that is, for the $i$th row, the decoder decides either $r_i = 0$ (i.e., $s_0$ has been sent through that row) or $r_i = 1$ (i.e., $s_1$ has been sent through that row). Assuming that message $m$ is sent, a decoding error occurs if either $r_m = 0$, or for any of $i \neq m$, $r_i = 1$. According to the Stein-Chernoff Lemma, for any fixed probability of $\mathrm{Pr}[r_m = 0|m \; \mathrm{sent}] = \epsilon$ and any fixed $\delta > 0$, we can achieve
\begin{eqnarray}
\mathrm{Pr}[r_i = 1|m \; \mathrm{sent}] \leq \exp\left\{-N [D(p_1\|p_0) - \delta]\right\},
\end{eqnarray}
as $N$ grows sufficiently large, for each $i \neq m$. Hence from the union bound, the decoding error probability with message $m$ sent is upper bounded by
\begin{eqnarray}
\label{eqn:union-bound-ppm}
&&\mathrm{Pr}[\mathrm{some}\; m^\prime \neq m \;\mathrm{declared}|m \;\mathrm{sent}]\nonumber\\
 &\leq& \epsilon + M \cdot\exp\left\{-N [D(p_1\|p_0) - \delta]\right\}.
\end{eqnarray}
So by choosing the coding rate appropriately as long as $(\log M)/N < D(p_1\|p_1) - \delta$, and then by letting $\epsilon$ and $\delta$ approach zero, the decoding error probability can be made arbitrarily close to zero. Noting that the total cost of transmitting the $M \times N$ matrix codeword is $N$, the achieved rate per unit cost thus can be made arbitrarily close to $\mathbf{C} = D(p_1\|p_1)$.

\begin{figure*}
\centering
\includegraphics[width=5in]{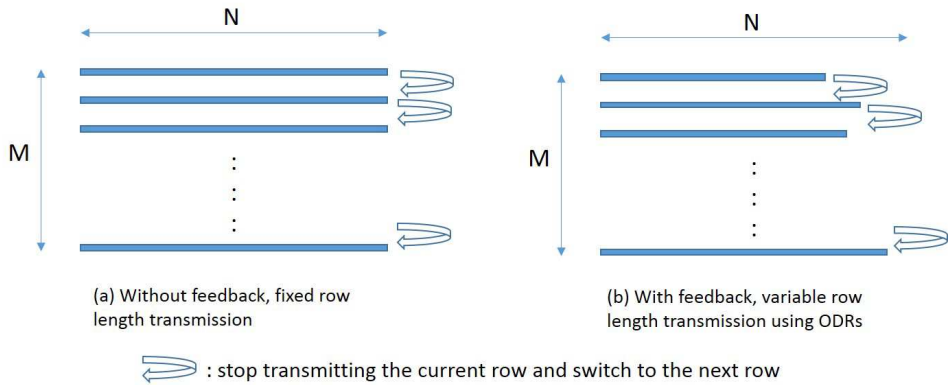}
\centering
\caption{Illustration of the transmission schemes without and with feedback.}
\label{fig:cap-uc-ppm}
\end{figure*}

When feedback is available, while receiving each row of the codeword matrix, the decoder can operate an ODR, so as to permit early termination with $p_1$ declared. Hence we can have an adaptive transmission scheme as follows (see Figure \ref{fig:cap-uc-ppm} for an illustration),
\begin{enumerate}
\item Set $i = 1$.
\item The encoder transmits the elements of the $i$th row of the corresponding codeword matrix in the root codebook, one by one; meanwhile the decoder performs an ODR with maximum sample size $N$.
\item The decoder informs the encoder through the feedback link its decision immediately when the decision is made using the ODR.
\item The encoder stops transmitting its current row once receiving the decision from the decoder, increases $i$ by one (unless $i = M$ already), and goes to Step 2).
\item If $i = M$, then the encoder halts; the decoder declares the decoded message to be $\hat{m}$ if there is only one row index $\hat{m}$ whose ODR detects $p_1$, and for all other cases (no such $\hat{m}$ exists or more than one such $\hat{m}$ exist) the decoder arbitrarily makes a declaration of the decoded message.
\end{enumerate}

We note that the above adaptive transmission scheme is feasible due to the availability of feedback, because then the encoder and the decoder can maintain synchronization regarding which row is being sent across the channel, even under ODR with variable stopping times.

Now let us suppose that there exist ODRs that outperform the asymptotic performance in Corollary \ref{cor:odr-exponent-falsealarm}; that is, at least for certain $0 < \eta < 1$, for an arbitrarily small but fixed target miss probability $P_\mathrm{M}^\ast > 0$, there exist ODRs such that when the normalized expected stopping time under $\mathcal{H}_1$, $T/N$, satisfies
\begin{eqnarray}
\label{eqn:supposed-odr-t}
\lim_{N \rightarrow \infty} \frac{T}{N} = \eta \leq 1,
\end{eqnarray}
the false alarm probability satisfies
\begin{eqnarray}
\label{eqn:supposed-odr-pfa}
\liminf_{N \rightarrow \infty} \frac{-\log P_\mathrm{FA}}{N} > \eta D(p_1\|p_0).
\end{eqnarray}

Let the decoder in the adaptive transmission scheme use the ODRs that we have supposed to exist satisfying (\ref{eqn:supposed-odr-t}) and (\ref{eqn:supposed-odr-pfa}). For such $\eta$, we have that the expected cost of sending a codeword is $\eta N + o(N)$ with $o(N)/N \rightarrow 0$ as $N \rightarrow \infty$, noting that transmitting all the rows, other than the one corresponding to the message index, incurs zero cost. On setting the probability $\mathrm{Pr}[r_m = 0|m \;\mathrm{sent}] = \epsilon > 0$ arbitrarily, according to (\ref{eqn:supposed-odr-pfa}) we have
\begin{eqnarray}
\mathrm{Pr}[r_i = 1| m \;\mathrm{sent}] \leq \exp\{-N \Delta\},
\end{eqnarray}
for some $\Delta > \eta D(p_1\|p_0)$, as $N$ grows without bound, for each $i \neq m$. Hence from the union bounding technique as that used in (\ref{eqn:union-bound-ppm}), the size of the message set can be made arbitrarily close to $\log M = N \Delta$, for achieving arbitrarily small decoding error probability as $N \rightarrow \infty$. Consequently, The achieved rate per unit cost is up to $N\Delta/(\eta N + o(N)) > N \eta D(p_1\|p_0)/(\eta N + o(N)) \rightarrow D(p_1\|p_0) = \mathbf{C}$, as $N \rightarrow \infty$. Therefore we encounter a contradiction since $\mathbf{C}$ cannot be outperformed by any coding scheme even in the presence of feedback and variable-length coding \cite{csiszar73:pcit}, and hence the supposed ODRs cannot exist. This establishes the converse part of Corollary \ref{cor:odr-exponent-falsealarm}.
$\Box$

\subsection{Completing the Proof of Theorem \ref{thm:odr-region}}
\label{subsec:completion}

Having established the two extremal cases in Corollaries \ref{cor:odr-exponent} and \ref{cor:odr-exponent-falsealarm}, in this subsection we complete the proof of Theorem \ref{thm:odr-region}.

{\it Proof of the Converse Part:} We prove that no ODRs may outperform the performance region in Theorem \ref{thm:odr-region}. First, note that for any $0 \leq \eta \leq 1$, the pair $(\Delta_\mathrm{FA}, \Delta_\mathrm{M})$ has to be bounded by
\begin{eqnarray}
\label{eqn:converse-fss}
\Delta_\mathrm{FA} &\leq& \sup_{\alpha > 0} \left\{\alpha \left[D(p_1\|p_0) - \nu [D(p_0\|p_1) + D(p_1\|p_0)]\right]\right.\nonumber\\
 &&\left.- \log \mathbb{E}_0\left[e^{\alpha \log p_1(\rvx)/p_0(\rvx)}\right]
\right\},\nonumber\\
\Delta_\mathrm{M} &\leq& \sup_{\alpha < 0} \left\{\alpha \left[D(p_1\|p_0) - \nu [D(p_0\|p_1) + D(p_1\|p_0)]\right]\right.\nonumber\\
 &&\left.- \log \mathbb{E}_1\left[e^{\alpha \log p_1(\rvx)/p_0(\rvx)}\right]
\right\},
\end{eqnarray}
for $0 \leq \nu \leq 1$, because that is the performance boundary achieved by the FSS likelihood ratio threshold test rules, under sample size $N$.

Second, according to Corollary \ref{cor:odr-exponent-falsealarm}, for any $0 \leq \eta \leq 1$, $\Delta_\mathrm{FA}$ has to satisfy
\begin{eqnarray}
\label{eqn:converse-fa}
\Delta_\mathrm{FA} \leq \eta D(p_1\|p_0).
\end{eqnarray}

The converse part of Theorem \ref{thm:odr-region} thus follows from combining (\ref{eqn:converse-fss}) and (\ref{eqn:converse-fa}).

{\it Proof of the Achievability Part:} We prove that ODRs exist attaining the performance region in Theorem \ref{thm:odr-region}. For this, using $2$-ODRs $\left\{\hat{\mathcal{H}}_2(M, \tau_M, \tau_N)\right\}$ with parameters
\begin{eqnarray}
M &=& \eta N\nonumber\\
\frac{\tau_M}{M} &=& D(p_1\|p_0) - \mu [D(p_0\|p_1) + D(p_1\|p_0)]\nonumber\\
\frac{\tau_N}{N} &=& D(p_1\|p_0) - \nu [D(p_0\|p_1) + D(p_1\|p_0)],
\end{eqnarray}
for $0 < \mu, \nu < 1$, we have the following achievable performance tuple:
\begin{eqnarray}
\label{eqn:region-tuple}
&&\Delta_\mathrm{FA} =\nonumber\\
&& \min\left\{
\sup_{\alpha > 0} \left\{\alpha \eta \left[D(p_1\|p_0) - \mu [D(p_0\|p_1) + D(p_1\|p_0)]\right]\right.\right.\nonumber\\
 &&\left.\left.- \eta \log \mathbb{E}_0\left[e^{\alpha \log p_1(\rvx)/p_0(\rvx)}\right]\right\},\right.\nonumber\\
&&\quad \left.
\sup_{\alpha > 0} \left\{\alpha \left[D(p_1\|p_0) - \nu [D(p_0\|p_1) + D(p_1\|p_0)]\right]\right.\right.\nonumber\\
 &&\left.\left.- \log \mathbb{E}_0\left[e^{\alpha \log p_1(\rvx)/p_0(\rvx)}\right]
\right\}
\right\}\nonumber\\
&&\Delta_\mathrm{M} =\nonumber\\
&& \max\left\{
\sup_{\alpha < 0} \left\{\alpha \eta \left[D(p_1\|p_0) - \mu [D(p_0\|p_1) + D(p_1\|p_0)]\right]\right.\right.\nonumber\\
 &&\left.\left.- \eta \log \mathbb{E}_1\left[e^{\alpha \log p_1(\rvx)/p_0(\rvx)}\right]\right\},\right.\nonumber\\
&&\quad \left.
\sup_{\alpha < 0} \left\{\alpha \left[D(p_1\|p_0) - \nu [D(p_0\|p_1) + D(p_1\|p_0)]\right]\right.\right.\nonumber\\
 &&\left.\left.- \log \mathbb{E}_1\left[e^{\alpha \log p_1(\rvx)/p_0(\rvx)}\right]
\right\}
\right\},
\end{eqnarray}
and $\lim_{N \rightarrow \infty} T/N = \eta$.

We need to prove that the above region (\ref{eqn:region-tuple}) contains the region described in Theorem \ref{thm:odr-region}. For any fixed $0 \leq \eta \leq 1$, denote the value of $\nu$ that solves
\begin{eqnarray}
\sup_{\alpha > 0} \left\{\alpha \left[D(p_1\|p_0) - \nu [D(p_0\|p_1) + D(p_1\|p_0)]\right] -\right.\nonumber\\
 \left.\log \mathbb{E}_0\left[e^{\alpha \log p_1(\rvx)/p_0(\rvx)}\right]
\right\} = \eta D(p_1\|p_0)
\end{eqnarray}
by $\nu^\ast$. What needs to be proved then is that, for any $\nu \geq \nu^\ast$, there exists a $\mu$ such that
\begin{eqnarray}
\label{eqn:containment-pre}
&&\eta \cdot \sup_{\alpha > 0} \left\{\alpha \left[D(p_1\|p_0) - \mu [D(p_0\|p_1) + D(p_1\|p_0)]\right]\right.\nonumber\\
&&\left. - \log \mathbb{E}_0\left[e^{\alpha \log p_1(\rvx)/p_0(\rvx)}\right]\right\} \geq\nonumber\\
&&\sup_{\alpha > 0} \left\{\alpha \left[D(p_1\|p_0) - \nu [D(p_0\|p_1) + D(p_1\|p_0)]\right]\right.\nonumber\\
&&\left. - \log \mathbb{E}_0\left[e^{\alpha \log p_1(\rvx)/p_0(\rvx)}\right]
\right\}, \;\mbox{and}\nonumber\\
&&\eta \cdot \sup_{\alpha < 0} \left\{\alpha \left[D(p_1\|p_0) - \mu [D(p_0\|p_1) + D(p_1\|p_0)]\right]\right.\nonumber\\
&&\left. - \log \mathbb{E}_1\left[e^{\alpha \log p_1(\rvx)/p_0(\rvx)}\right]\right\} \leq\nonumber\\
&&\sup_{\alpha < 0} \left\{\alpha \left[D(p_1\|p_0) - \nu [D(p_0\|p_1) + D(p_1\|p_0)]\right]\right.\nonumber\\
&&\left. - \log \mathbb{E}_1\left[e^{\alpha \log p_1(\rvx)/p_0(\rvx)}\right]
\right\}.
\end{eqnarray}
This is because, if the two inequalities in (\ref{eqn:containment-pre}) holds, then we have that the tuple
\begin{eqnarray}
\Delta_\mathrm{FA} &=& \sup_{\alpha > 0} \left\{\alpha \left[D(p_1\|p_0) - \nu [D(p_0\|p_1) + D(p_1\|p_0)]\right]\right.\nonumber\\
&&\left. - \log \mathbb{E}_0\left[e^{\alpha \log p_1(\rvx)/p_0(\rvx)}\right]
\right\}\nonumber\\
\Delta_\mathrm{M} &=& \sup_{\alpha < 0} \left\{\alpha \left[D(p_1\|p_0) - \nu [D(p_0\|p_1) + D(p_1\|p_0)]\right]\right.\nonumber\\
&&\left. - \log \mathbb{E}_1\left[e^{\alpha \log p_1(\rvx)/p_0(\rvx)}\right]
\right\}
\end{eqnarray}
is achievable, for any $\nu^\ast \leq \nu < 1$, and thus the region (\ref{eqn:region-tuple}) contains the region described in Theorem \ref{thm:odr-region}. But clearly letting $\mu$ be sufficiently close to zero suffices to satisfy (\ref{eqn:containment-pre}), because
\begin{eqnarray}
&&\eta \cdot \sup_{\alpha > 0} \left\{\alpha \left[D(p_1\|p_0) - 0 \cdot [D(p_0\|p_1) + D(p_1\|p_0)]\right]\right.\nonumber\\
&&\left. - \log \mathbb{E}_0\left[e^{\alpha \log p_1(\rvx)/p_0(\rvx)}\right]\right\} = \eta D(p_1\|p_0)\nonumber\\
&=& \sup_{\alpha > 0} \left\{\alpha \left[D(p_1\|p_0) - \nu^\ast \cdot [D(p_0\|p_1) + D(p_1\|p_0)]\right]\right.\nonumber\\
&&\left. - \log \mathbb{E}_0\left[e^{\alpha \log p_1(\rvx)/p_0(\rvx)}\right]\right\}\nonumber\\
&\geq& \sup_{\alpha > 0} \left\{\alpha \left[D(p_1\|p_0) - \nu \cdot [D(p_0\|p_1) + D(p_1\|p_0)]\right]\right.\nonumber\\
&&\left. - \log \mathbb{E}_0\left[e^{\alpha \log p_1(\rvx)/p_0(\rvx)}\right]\right\}
\end{eqnarray}
for any $\nu \geq \nu^\ast$, and
\begin{eqnarray}
&&\eta \cdot \sup_{\alpha < 0} \left\{\alpha \left[D(p_1\|p_0) - 0 \cdot [D(p_0\|p_1) + D(p_1\|p_0)]\right]\right.\nonumber\\
&&\left. - \log \mathbb{E}_1\left[e^{\alpha \log p_1(\rvx)/p_0(\rvx)}\right]\right\}\nonumber\\
&=& 0 \leq \sup_{\alpha < 0} \left\{\alpha \left[D(p_1\|p_0) - \nu [D(p_0\|p_1) + D(p_1\|p_0)]\right]\right.\nonumber\\
&&\left. - \log \mathbb{E}_1\left[e^{\alpha \log p_1(\rvx)/p_0(\rvx)}\right]
\right\}.
\end{eqnarray}
This thus completes the achievability part of Theorem \ref{thm:odr-region}.

{\it Discussion:} A pivotal operating point for FSS decision rules is that when
\begin{eqnarray}
&&\sup_{\alpha > 0} \left\{\alpha \left[D(p_1\|p_0) - \nu \cdot [D(p_0\|p_1) + D(p_1\|p_0)]\right]\right.\nonumber\\
&&\left. - \log \mathbb{E}_0\left[e^{\alpha \log p_1(\rvx)/p_0(\rvx)}\right]\right\}\nonumber\\
&=& \sup_{\alpha < 0} \left\{\alpha \left[D(p_1\|p_0) - \nu [D(p_0\|p_1) + D(p_1\|p_0)]\right]\right.\nonumber\\
&&\left. - \log \mathbb{E}_1\left[e^{\alpha \log p_1(\rvx)/p_0(\rvx)}\right]
\right\}
\end{eqnarray}
holds, and then their common value is exactly the Chernoff information of $(p_0, p_1)$, $C(p_0, p_1)$, and can be equivalently expressed as
\begin{eqnarray}
C(p_0, p_1) = -\inf_{\alpha \in (0, 1)} \log \left(\int_{\mathcal{X}} p_0^\alpha(x) p_1^{1 - \alpha}(x) dx\right),
\end{eqnarray}
an expression which has been used in Theorem \ref{thm:odr-fixed}. From the above proof of Theorem \ref{thm:odr-region}, an immediate consequence is that ODRs may achieve the operating point of $\Delta_\mathrm{FA} = \Delta_\mathrm{M} = C(p_0, p_1)$ if and only if $\eta \geq C(p_0, p_1)/D(p_1\|p_0)$.

\section{Conclusion}
\label{sec:conclusion}

In this paper, we have formulated the general ODR framework and treated several of its key characteristics. We have considered both finite and asymptotic problems. In the finite regime, we have established Bayesian optimal ODRs for the case of a fixed maximum sample size, and the case of a geometrically distributed maximum sample size. For the latter, the Bayesian optimal ODR is a likelihood ratio threshold test with two thresholds. In the asymptotic regime, as the maximum sample size grows without bound, we have completely characterized the tradeoff among the exponents of the (false alarm and miss) error probabilities and the normalized expected stopping time under the alternative hypothesis.

An interesting problem beyond the scope of this paper concerns the asymptotic analysis of the Bayesian optimal ODR. In such problems in the sequential analysis literature, one usually proceeds by letting the sampling cost $c$ decrease toward zero in the Bayesian cost; see, e.g., \cite[Sec. 13]{chernoff72:book}. For our setup, in order to make the problem meaningful, we need to further tune the growth of the maximum sample size $N$ (or the mean maximum sample size $1/\epsilon$ in the case of geometrically distributed maximum sample size) accordingly, say, following $O(1/c)$, and the interplay between the sampling cost and the maximum sample size may exhibit interesting behaviors.


\end{document}